\documentclass[%
reprint,
superscriptaddress,
 amsmath,amssymb,
 aps,
 pre,
]{revtex4-2}

\usepackage{graphicx}
\usepackage{dcolumn}
\usepackage{bm}
\usepackage[dvipsnames]{xcolor}
\usepackage{bbold}
\usepackage{mathtools}
\usepackage{xurl}
\usepackage{hyperref}

\begin{document}
\newcommand{\tr}{\text{Tr}}
\newcommand{\dpartial}{\dot{\partial}}
\newcommand{\bpartial}{\bar{\partial}}
\newcommand{\lagr}{\mathcal{L}}
\newcommand{\lagrgfm}{\mathcal{L}_{\text{GFM}}}
\newcommand{\lagrgf}{\mathcal{L}_{\text{GF}}}
\newcommand{\ham}{\mathcal{H}}
\newcommand{\hamgfm}{\mathcal{H}_{\text{GFM}}}
\newcommand{\hamgf}{\mathcal{H}_{\text{GF}}}
\newcommand{\attention}[1]{\textcolor{red}{#1}}
\newcommand{\ivan}[1]{\textcolor{red}{#1}}
\newcommand{\lit}{\textcolor{OliveGreen}{xxxADD CITATIONxxx }}
\newcommand{\morelit}{\textcolor{OliveGreen}{xxxADD MORE CITATIONSxxx }}
\newcommand{\secref}{\textcolor{purple}{xxxADD SECTIONxxx }}
\newcommand{\checkerr}{\textcolor{brown}{xxxCHECK EQUATIONSxxx }}
\newcommand{\leftbracket}{\bm{\{}\{}
\newcommand{\rightbracket}{\}\bm{\}}}
\newcommand{\PL}{\stackrel{\mathclap{\normalfont\mbox{pl}}}{=}}
\newcommand{\warning}[1]{\textcolor{Red}{#1}}

\title{Backpropagation as Dirac-bracket dynamics of dissipative training}

\author{N. L. Holanda}
\email{linneuholanda@gmail.com}
\affiliation{Independent researcher, S\~{a}o Paulo, Brazil}

\author{I. E. Cunha}
\email{ivan.cunha@udesc.br}
\affiliation{Department of Biological Sciences, Santa Catarina State University (UDESC), Rua Coronel Fernandes Martins 270, Laguna, Santa Catarina 88790-000, Brazil}

\date{\today}

\begin{abstract}
Gradient-based training of a parametric model is the motion of a damped particle in the potential set by the loss. Using the Bateman--Galley doubling of the degrees of freedom, which gives dissipative systems an action principle, we write gradient flow, the heavy-ball method and Nesterov's accelerated gradient method as Hamiltonian systems. We show that, within this Hamiltonian description, backpropagation is Dirac evolution. The activations of a feed-forward network are treated as coordinates of phase space, and the layer equations as constraints. These constraints are second class for every architecture, and Hamilton's equations with the resulting Dirac bracket reproduce training with the exact gradient of the loss, although the Hamiltonian contains only the cost of the network output. The Dirac bracket of an activation with the momenta of the parameters is its Jacobian, the inverse of the constraint matrix contains both the forward and the reverse modes of automatic differentiation, and the errors propagated by backpropagation are the Lagrange multipliers of the layer constraints: the constraint forces of a kinematic chain under the load of the cost. These results hold in the full doubled phase space, not only in the physical limit. The network's part of the constraint algebra is the same for every optimizer, Nesterov's look-ahead adds one layer of momentum-dependent constraints that couple the two copies of the doubled system, and gradient flow is obtained by constructing the Dirac bracket in stages. We confirm these statements numerically to machine precision and train a network by integrating the Dirac flow of its constraints, with no backward pass in the computation.
\end{abstract}

\maketitle

\section{Introduction}
\label{section_introduction}

Training a parametric model by gradient descent is a dynamical process: the parameters move in the potential defined by the loss, and the optimizer sets their inertia and friction. Momentum methods are discretizations of the damped motion of a particle, with the momentum parameter playing the role of a mass in a viscous medium~\cite{qian1999momentum}, and continuous-time models of this kind have clarified the behavior of momentum and acceleration~\cite{su2016differential, shi2022understanding, kovachki2021continuous}. A Lagrangian or Hamiltonian description of training would make the tools of analytical mechanics available to study it, but optimizers dissipate energy, and Hamilton's principle in its usual form does not apply. Accelerated methods have been generated from explicitly time-dependent Lagrangians~\cite{wibisono2016variational}, studied as dissipative Hamiltonian systems~\cite{diakonikolas2021generalized}, and discretized with symplectic and conformal symplectic integrators~\cite{betancourt2018symplectic, franca2020conformal}, and Lagrangian descriptions of learning have been used to derive the broken conservation laws of training~\cite{kunin2020neural, tanaka2021noether}.

Here we use the doubling of the degrees of freedom introduced by Bateman~\cite{bateman1931dissipative} and generalized by Galley~\cite{galley2013classical, galley2014principle}. The doubled system has an ordinary action and Hamiltonian, and the physical motion is recovered when the two copies are identified after the equations of motion have been derived. Gradient flow, the heavy-ball method~\cite{polyak1964some} and Nesterov's accelerated gradient method~\cite{nesterov1983method} all take this form, with the learning rate and the momentum coefficient set by the mass, the damping and the time step.

The question we address is how the gradient of the loss arises in this Hamiltonian. For a neural network, the gradient is computed by backpropagation~\cite{rumelhart1986learning}, and a Hamiltonian that simply contains the loss as a potential takes the gradient as given. We show instead that backpropagation is part of the Hamiltonian dynamics. We treat the activations of a feed-forward network as additional coordinates of phase space, give them no dynamics of their own, and impose the layer equations as holonomic constraints. The Hamiltonian then contains only the cost of the network output. The layer constraints are second class for every architecture, and Dirac's theory of constrained systems~\cite{dirac1950generalized, henneaux1992quantization} gives a bracket in which Hamilton's equations reproduce training with the exact gradient of the loss. The inverse of the constraint matrix is block triangular: one of its blocks propagates tangents forward through the network, which is forward-mode differentiation, and the other propagates errors backward, which is backpropagation. The Dirac bracket of an activation with the momenta of the parameters is its Jacobian, and the errors are the Lagrange multipliers of the layer constraints. Mechanically, the network is a kinematic chain, such as a robot arm, and the backward pass is the statics of this chain: the computation of the constraint forces carried by its links under the load of the cost.

The identification of the errors of backpropagation with Lagrange multipliers is old. In the Lagrangian formulation of backpropagation~\cite{lecun1988theoretical}, as in the adjoint method used to train continuous-depth networks~\cite{chen2018neural}, the multipliers enforce the layer equations along the depth of the network. More recently, backpropagation has been derived from a Bateman--Galley action in which the depth of the network plays the role of time and the doubled variables encode the activations and the errors~\cite{scurria2026physical}. Our construction differs from both: the time is the training time, the network enters as a set of constraints on the training dynamics, and backpropagation follows from the Dirac bracket of the optimizer itself.

The fact that a Dirac bracket reduces to the bracket of the constraint surface is general, and it guarantees that the Dirac evolution contains the gradient of the loss. Our contributions are the explicit structure of this reduction for neural networks and its consequences for training. (i) The network's part of the constraint algebra, its inverse and the resulting Dirac bracket are the same for every optimizer described by the Bateman--Galley Hamiltonian, with arbitrary time-dependent mass and damping. (ii) All statements hold in the full doubled phase space, not only in the physical limit. (iii) Nesterov's method enters through one additional layer of constraints that depend on the momenta, so that the backward pass is evaluated at the look-ahead point. Unlike the layer constraints, these constraints have nonvanishing brackets between the two copies of the doubled system, which they couple; the effect of this coupling on the equations of motion vanishes in the physical limit. (iv) Gradient flow, whose own Hamiltonian description already requires Dirac brackets, is obtained by constructing the Dirac bracket in stages. (v) A network can be trained by integrating the Dirac flow of its constraints, with no backward pass in the computation, which we verify numerically.

The paper is organized as follows. Section~\ref{section_hamiltonian_learning} summarizes the Bateman--Galley Hamiltonian of the optimizers. Section~\ref{section_backpropagation} derives backpropagation from the Dirac bracket of the layer constraints, and Sec.~\ref{section_numerical_test} tests it numerically. We conclude in Sec.~\ref{section_conclusion}. Appendix~\ref{appendix_hamiltonian_mechanics_gradient_flow} treats gradient flow, Appendix~\ref{appendix_two_layer_example} works out a network with one hidden layer, and Appendix~\ref{appendix_nesterov_lookahead} derives the look-ahead constraints of Nesterov's method.

\section{Hamiltonian mechanics of dissipative training}
\label{section_hamiltonian_learning}

In this section we summarize the Bateman--Galley description of gradient-based optimizers on which the rest of the paper is built.

\subsection{Optimizers as damped motion}
The continuous-time dynamics of the optimizers we consider is the damped motion of a particle,
\begin{equation}
\label{ode}
\mu(t)\frac{d^2\mathbf{x}}{dt^2} + \gamma (t)\frac{d\mathbf{x}}{dt} + \nabla L(\mathbf{x}) = 0,
\end{equation}
where $\mathbf{x}$ collects the parameters of the model, the loss $L(\mathbf{x})$ is the potential, and $\mu(t)$ and $\gamma(t)$ are the mass and the damping coefficient, which may depend on time. For $\mu = 0$, equation \eqref{ode} is gradient flow,
\begin{equation}
\label{gradient_flow}
\frac{d\mathbf{x}}{dt} = -\frac{1}{\gamma(t)}\nabla_{\mathbf{x}} L,
\end{equation}
whose discretization on time steps of length $h$ is gradient descent with learning rate $\lambda = h/\gamma$,
\begin{equation}
\label{discrete_gradient_flow_damping}
\mathbf{x}[n+1] = \mathbf{x}[n] - \frac{h}{\gamma(t)}\nabla L(\mathbf{x}[n]).
\end{equation}
For $\mu \neq 0$, the central difference for the acceleration and the backward difference for the velocity give Polyak's heavy-ball method~\cite{polyak1964some},
\begin{equation}
\label{polyak_heavy_ball}
\mathbf{x}[n+1] = \mathbf{x}[n] + \beta\big(\mathbf{x}[n] - \mathbf{x}[n-1]\big) - \alpha\,\nabla L\big(\mathbf{x}[n]\big),
\end{equation}
with learning rate $\alpha = h^2/\mu$ and momentum coefficient $\beta = 1 - h\gamma/\mu$. Nesterov's accelerated gradient method~\cite{nesterov1983method} evaluates the gradient at the extrapolated point $\tilde{\mathbf{x}}[n] = \mathbf{x}[n] + \beta\big(\mathbf{x}[n] - \mathbf{x}[n-1]\big)$ instead,
\begin{equation}
\label{nesterov}
\mathbf{x}[n+1] = \tilde{\mathbf{x}}[n] - \alpha\,\nabla L\big(\tilde{\mathbf{x}}[n]\big).
\end{equation}
Through first order in $h$, this adds the Hessian-driven damping $h\,\nabla\nabla L(\mathbf{x})\,\dot{\mathbf{x}}$ to equation \eqref{ode}~\cite{alvarez2002second, shi2022understanding}, and its common schedule $\beta[n] = (n-1)/(n+2)$ corresponds to $\mu = 1$ and $\gamma(t) = 3/t$~\cite{su2016differential}. We keep $\mu$ and $\gamma$ time dependent throughout.

\subsection{Bateman--Galley doubling}
Equation \eqref{ode} is dissipative and does not follow from an ordinary action. Following Bateman~\cite{bateman1931dissipative} and Galley~\cite{galley2013classical, galley2014principle}, we double the degrees of freedom, $\mathbf{x} \to (\mathbf{x}_1, \mathbf{x}_2)$, and write the two copies as a Minkowski vector. With latin indices $a, b$ running over the $2D$ components, the contravariant and covariant vectors are $x^a = (\mathbf{x}_1, \mathbf{x}_2)$ and $x_a = \eta_{ab}x^b = (\mathbf{x}_1, -\mathbf{x}_2)$, with the metric $\eta = \mathrm{diag}(\mathbb{I}_D, -\mathbb{I}_D)$ and the summation convention, so that $x^ap_a = \mathbf{x}_1\cdot\mathbf{p}_1 - \mathbf{x}_2\cdot\mathbf{p}_2$. The matrix
\begin{equation}
\label{cal_x}
X = \begin{bmatrix}
0 & \mathbb{I}_D\\
\mathbb{I}_D & 0
\end{bmatrix}
\end{equation}
exchanges the copies and defines the Bateman--Galley conjugate ${x^\dagger}^a = X^a{}_b x^b$, with $p^a x^\dagger_a = \mathbf{p}_1 \cdot \mathbf{x}_2 - \mathbf{p}_2 \cdot \mathbf{x}_1$. The loss enters through the antisymmetric potential $V = L(\mathbf{x}_1) - L(\mathbf{x}_2)$, and derivatives are written as $\partial^a = \partial/\partial x_a = (\partial/\partial\mathbf{x}_1, -\partial/\partial\mathbf{x}_2)$ and, for the velocities, $\dpartial^a = (\partial/\partial\dot{\mathbf{x}}_1, -\partial/\partial\dot{\mathbf{x}}_2)$.

The Lagrangian of equation \eqref{ode} is
\begin{align}
\label{lagrangian}
\lagr =& -L(\mathbf{x}_1) + L(\mathbf{x}_2) + \frac{1}{4}\big[\dot{\gamma}-\ddot{\mu}\big](\mathbf{x}_1^2 - \mathbf{x}_2^2) \nonumber\\
&+ \frac{1}{2}\big[\gamma-\dot{\mu}\big](\dot{\mathbf{x}}_1\cdot\mathbf{x}_2 - \dot{\mathbf{x}}_2\cdot\mathbf{x}_1) + \frac{1}{2}\mu(\dot{\mathbf{x}}_1^2 - \dot{\mathbf{x}}_2^2),
\end{align}
or, in covariant form,
\begin{equation}
\label{lagrangian_cov}
\lagr = -V + \frac{1}{4}\big[\dot{\gamma} - \ddot{\mu}\big]x^ax_a + \frac{1}{2}\big[\gamma - \dot{\mu}\big]\dot{x}^a x^\dagger_a + \frac{1}{2}\mu\,\dot{x}^a\dot{x}_a,
\end{equation}
where the time dependence of $\mu$ and $\gamma$ is implicit. The action $\mathcal{S} = \int\lagr\,dt$ is stationary on the solutions of the Euler--Lagrange equations
\begin{equation}
\label{euler_lagrange_eq_cov}
\partial^a\lagr - \frac{d}{dt}\dpartial^a\lagr = 0.
\end{equation}
These equations hold exactly for the doubled system. The submanifold $\mathbf{x}_1 = \mathbf{x}_2$, $\dot{\mathbf{x}}_1 = \dot{\mathbf{x}}_2$ is invariant under the doubled dynamics, and on it both copies obey equation \eqref{ode}. Identifying the copies after all derivatives have been taken is the \emph{physical limit}, and we write $\PL$ for equalities that hold only there. The physical limit is invariant but unstable: the second copy, which absorbs the energy dissipated by the first, is anti-damped, so that deviations of the two copies from each other grow.

\subsection{Hamiltonian and brackets}
The conjugate momenta are $p^a = \dpartial^a\lagr = (\mathbf{p}_1, \mathbf{p}_2)$, and we write $\bpartial_a = \partial/\partial p^a$. The Legendre transformation
\begin{equation}
\label{legendre_transformation_cov}
\ham = p^a\dot{x}_a - \lagr
\end{equation}
gives the Hamiltonian
\begin{equation}
\label{hamiltonian_cov}
\mathcal{H} =   \frac{1}{2\mu}p^ap_a - \frac{\gamma - \dot{\mu}}{2\mu}p^ax^\dagger_a - \Big[\frac{\dot{\gamma} - \ddot{\mu}}{4} + \frac{(\gamma - \dot{\mu})^2}{8\mu}\Big]x^ax_a + V,
\end{equation}
and Hamilton's equations
\begin{equation}
\label{hamilton_equations_cov}
\frac{dx^a}{dt} = \bpartial^a \mathcal{H}, \qquad
\frac{dp^a}{dt} = - \partial^a \mathcal{H},
\end{equation}
which hold exactly in the doubled phase space. On the invariant submanifold $\mathbf{x}_1 = \mathbf{x}_2$, $\mathbf{p}_1 = \mathbf{p}_2$, both copies obey
\begin{subequations}
\label{hamilton_equations_euc}
\begin{equation}
\frac{d\mathbf{x}}{dt} = \frac{1}{\mu}\mathbf{p} - \frac{\gamma - \dot{\mu}}{2\mu}\mathbf{x},
\end{equation}
\begin{equation}
\frac{d\mathbf{p}}{dt} = -\nabla L(\mathbf{x}) - \frac{\gamma - \dot{\mu}}{2\mu}\mathbf{p} + \bigg( \frac{\dot{\gamma} - \ddot{\mu}}{2} + \frac{(\gamma - \dot{\mu})^2}{4\mu}\bigg)\mathbf{x},
\end{equation}
\end{subequations}
which are equivalent to equation \eqref{ode}.

The Poisson bracket of two phase-space functions is
\begin{equation}
\label{scalar_poisson_bracket}
\Gamma(f, g) = \frac{\partial f}{\partial x_a}\frac{\partial g}{\partial p^a} - \frac{\partial f}{\partial p_a}\frac{\partial g}{\partial x^a} = \Gamma_1(f, g) - \Gamma_2(f, g),
\end{equation}
where $\Gamma_i$ is the canonical bracket in the variables $(\mathbf{x}_i, \mathbf{p}_i)$ of copy $i$, and every phase-space function evolves as
\begin{equation}
\label{poisson_em}
\frac{df}{dt} = \Gamma(f, \mathcal{H}) + \frac{\partial f}{\partial t}.
\end{equation}
For gradient flow, $\mu = 0$, the Lagrangian is linear in the velocities, the momenta are fixed functions of the coordinates, and the Legendre transformation is singular. The resulting constraints are second class, and the Hamiltonian description requires Dirac brackets~\cite{dirac1950generalized, gitman1990quantization, henneaux1992quantization}, as shown in Appendix~\ref{appendix_hamiltonian_mechanics_gradient_flow}.

\subsection{Nesterov's look-ahead}
\label{section_nesterov_lookahead}
The Hessian-driven damping of Nesterov's method can be included in the doubled Lagrangian by adding the term $\frac{h}{2}\big(\dot{\mathbf{x}}_1\cdot\nabla L(\mathbf{x}_2) - \dot{\mathbf{x}}_2\cdot\nabla L(\mathbf{x}_1)\big)$ to equation \eqref{lagrangian}, but the resulting Hamiltonian contains the gradient of the loss explicitly. An equivalent description, closer to the algorithm, evaluates the loss at a look-ahead point. Let $\ham_0$ be the Hamiltonian \eqref{hamiltonian_cov} without the potential $V$, let $v^a = \bpartial^a\ham_0$ be the velocities of the two copies as functions of the canonical coordinates, and let $\bar{\mathbf{v}} = (\mathbf{v}_1 + \mathbf{v}_2)/2$ be their mean. The Hamiltonian
\begin{equation}
\label{nesterov_lookahead_hamiltonian}
\ham_{\text{LA}} = \ham_0 + L\big(\mathbf{x}_1 + h\bar{\mathbf{v}}\big) - L\big(\mathbf{x}_2 + h\bar{\mathbf{v}}\big)
\end{equation}
depends on the loss only through its values at the look-ahead points $\mathbf{x}_i + h\bar{\mathbf{v}}$. In the physical limit, the contributions of the look-ahead to the velocities cancel between the two copies, while the force on the momenta is the gradient at the look-ahead point, and Hamilton's equations give
\begin{equation}
\label{nesterov_lookahead_ode}
\mu\frac{d^2\mathbf{x}}{dt^2} + \gamma\frac{d\mathbf{x}}{dt} + \nabla L\Big(\mathbf{x} + h\frac{d\mathbf{x}}{dt}\Big) = 0,
\end{equation}
the continuum counterpart of the gradient step \eqref{nesterov} from the extrapolated point. Expanding the gradient to first order in $h$ recovers the Hessian-driven damping.

\section{Backpropagation as Dirac evolution}
\label{section_backpropagation}
In this section we show that backpropagation is generated by the Dirac bracket of the Hamiltonian description of Sec.~\ref{section_hamiltonian_learning}.

The idea is to treat the activations of a neural network as additional coordinates of phase space, tied to the parameters by the equations that define the network. These equations are second-class constraints, and the Dirac bracket that enforces them generates, through the inverse of the constraint matrix, both the forward and the backward pass of the network. In this picture, the Hamiltonian contains the loss only as a function of the network output, and the gradient of the loss with respect to the parameters is never written down explicitly: it emerges from the constraint algebra.

\subsection{Data, model and loss}
The data set consists of $N$ input--target pairs $(\mathbf{h}_0^{(n)}, \mathbf{y}^{(n)})$, $n = 1, \dots, N$. A feed-forward network with $L$ layers maps each input through the activations
\begin{equation}
\label{layer_map}
\mathbf{h}_\ell^{(n)} = \mathbf{f}_\ell\big(W_\ell, \mathbf{h}_{\ell-1}^{(n)}\big), \qquad \ell = 1, \dots, L,
\end{equation}
where $W_\ell$ are the parameters of layer $\ell$; for a dense layer, $\mathbf{f}_\ell(W_\ell, \mathbf{h}) = \sigma(W_\ell\mathbf{h})$. The output $\mathbf{h}_L^{(n)}$ is scored by a cost $c^{(n)}(\mathbf{h}_L^{(n)})$, which also depends on the target $\mathbf{y}^{(n)}$, for example the squared error \eqref{two_layer_least_squares} for regression or the cross-entropy \eqref{two_layer_cross_entropy} for classification. The vector of parameters $\mathbf{x}$ collects the entries of all $W_\ell$, and the loss is the mean cost over the data set,
\begin{equation}
\label{network_loss}
L(\mathbf{x}) = \frac{1}{N}\sum_{n=1}^N c^{(n)}\big(\mathbf{h}_L^{(n)}(\mathbf{x})\big).
\end{equation}
The linear model of least-squares regression is the case $L = 1$ with $\mathbf{f}_1(W_1, \mathbf{h}) = W_1\mathbf{h}$.

Training this network with any of the optimizers of Sec.~\ref{section_hamiltonian_learning} means evolving the doubled parameters with the Hamiltonian \eqref{hamiltonian_cov} and the potential $V = L(\mathbf{x}_1) - L(\mathbf{x}_2)$. Hamilton's equations \eqref{hamilton_equations_cov} then require the force $\partial^a V = \big(\nabla L(\mathbf{x}_1), \nabla L(\mathbf{x}_2)\big)$, that is, the gradient of the loss with respect to every parameter of the network, which in practice is computed by backpropagation. In the rest of this section we show that this gradient need not be supplied by hand: it follows from a Hamiltonian that depends only on the cost of the network output, once the layers are treated as constraints. We denote the Jacobians of the layer maps by
\begin{equation}
\label{layer_jacobians}
J_\ell = \frac{\partial \mathbf{f}_\ell}{\partial \mathbf{h}_{\ell-1}}, \qquad G_\ell = \frac{\partial \mathbf{f}_\ell}{\partial W_\ell}.
\end{equation}

\subsection{Layers as constraints}
We enlarge the configuration space of Sec.~\ref{section_hamiltonian_learning}. In addition to the doubled parameters $\mathbf{x}_1, \mathbf{x}_2$, each copy $i = 1, 2$ of the system carries the activations $\mathbf{h}_{\ell, i}$ of every layer. To keep the notation light, $\mathbf{h}_\ell$ denotes the activations of layer $\ell$ for all samples of the data set, stacked into a single vector; the layer maps \eqref{layer_map} act sample by sample, so that their Jacobians \eqref{layer_jacobians} are block diagonal over samples, and $c(\mathbf{h}_L) = \frac{1}{N}\sum_n c^{(n)}(\mathbf{h}_L^{(n)})$ is the cost of the whole data set. The activations carry no kinetic or damping terms, and the Lagrangian depends on them only through the cost of the network output,
\begin{equation}
\label{network_lagrangian}
\lagr = \lagr_0(x, \dot{x}, t) - c(\mathbf{h}_{L,1}) + c(\mathbf{h}_{L,2}),
\end{equation}
where $\lagr_0$ is the Lagrangian \eqref{lagrangian_cov} without the potential $-V$. Since the activations have no velocities in the Lagrangian, their conjugate momenta $\bm{\pi}_{\ell, i}$ vanish, and the Legendre transformation \eqref{legendre_transformation_cov} gives
\begin{equation}
\label{network_hamiltonian}
\mathcal{H} = \mathcal{H}_0 + c(\mathbf{h}_{L,1}) - c(\mathbf{h}_{L,2}),
\end{equation}
where $\mathcal{H}_0$ is the Hamiltonian \eqref{hamiltonian_cov} without the potential $V$. The Hamiltonian \eqref{network_hamiltonian} contains no derivative of the loss with respect to the parameters; the network itself enters only through the layer equations \eqref{layer_map}, which we impose as holonomic constraints on the enlarged configuration space. Together with the vanishing momenta, they give the constraints
\begin{equation}
\label{layer_constraints}
\bm{\phi}_{\ell, i} = \mathbf{h}_{\ell, i} - \mathbf{f}_\ell(W_{\ell, i}, \mathbf{h}_{\ell-1, i}) \approx 0, \qquad \bm{\pi}_{\ell, i} \approx 0,
\end{equation}
where $\approx$ denotes Dirac's weak equality, that is, equality on the constraint surface.

The layers of the network are therefore constraints, and the activations are coordinates without dynamics of their own, rigidly slaved to the parameters. Mechanically, the network is a kinematic chain, such as a robot arm: the parameters are the joint variables, the activations are the positions of the successive links, each layer equation is the rigid connection between two consecutive links, and the output is the end effector, on which the cost acts as the potential energy of an external load. We return to this picture after deriving the equations of motion.

The brackets between constraints of different copies vanish. Because $\Gamma = \Gamma_1 - \Gamma_2$ and the cost enters the Hamiltonian \eqref{network_hamiltonian} with opposite signs for the two copies, the two signs cancel, and both copies obey identical equations. In the remainder of this section we therefore drop the copy index $i$ and work with $\Gamma_1$.

\subsection{The constraint algebra}
\label{section_constraint_algebra}
We first deal with the general case, allowing the constraints to depend explicitly on time. Let $\Theta \approx 0$ be a set of second-class constraints, so that the matrix of their brackets, $\mathcal{M} = \Gamma(\Theta, \Theta)$, is invertible. The constrained evolution is generated by the total Hamiltonian $\mathcal{H}_T = \mathcal{H} + u\cdot\Theta$,
\begin{equation}
\label{total_evolution}
\frac{df}{dt} = \Gamma(f, \mathcal{H}) + \Gamma(f, \Theta)\,u + \frac{\partial f}{\partial t},
\end{equation}
with the multipliers $u$ fixed by the requirement that the constraints be preserved in time, $\Gamma(\Theta, \mathcal{H}) + \mathcal{M}\,u + \partial\Theta/\partial t = 0$, that is, $u = -\mathcal{M}^{-1}\big[\Gamma(\Theta, \mathcal{H}) + \partial\Theta/\partial t\big]$. Eliminating them, we find
\begin{equation}
\label{dirac_evolution_time_dependent}
\frac{df}{dt} = \Delta(f, \mathcal{H}) + \frac{\partial f}{\partial t} - \Gamma(f, \Theta)\,\mathcal{M}^{-1}\,\frac{\partial\Theta}{\partial t},
\end{equation}
where $\Delta(f, g) = \Gamma(f, g) - \Gamma(f, \Theta)\,\mathcal{M}^{-1}\,\Gamma(\Theta, g)$ is the Dirac bracket. Both the layer constraints \eqref{layer_constraints} and the look-ahead constraints of Nesterov's method (Appendix~\ref{appendix_nesterov_lookahead}) are of two kinds: constraints $P \approx 0$ that set momenta to zero, and constraints $Q \approx 0$ that do not contain these momenta. Thus $\Gamma(P, P) = 0$ and, with $C \equiv \Gamma(Q, Q)$ and $D \equiv \Gamma(Q, P)$,
\begin{equation}
\label{constraint_matrix_blocks}
\begin{gathered}
\mathcal{M} = \begin{bmatrix} C & D\\ -D^{T} & 0 \end{bmatrix},\\[2pt]
\mathcal{M}^{-1} = \begin{bmatrix} 0 & -D^{-T}\\ D^{-1} & D^{-1}C\,D^{-T} \end{bmatrix}.
\end{gathered}
\end{equation}
Since $P \approx 0$ does not depend on time, only $-\Gamma(f, P)\,D^{-1}\,\partial Q/\partial t$ remains of the last term of equation \eqref{dirac_evolution_time_dependent}, so that an explicit time dependence of the constraints affects only the functions that have nonvanishing brackets with $P$.

For the layer constraints, $Q = \bm{\phi}$ and $P = \bm{\pi}$. The constraints $\bm{\phi}_\ell$ contain no momenta, and the constraints $\bm{\pi}_\ell$ contain no coordinates, so that $\Gamma(\bm{\phi}_\ell, \bm{\phi}_k) = \Gamma(\bm{\pi}_\ell, \bm{\pi}_k) = 0$, and $C = 0$. The only nonvanishing brackets are
\begin{equation}
\label{constraint_brackets_network}
\Gamma(\bm{\phi}_\ell, \bm{\pi}_k) = \frac{\partial \bm{\phi}_\ell}{\partial \mathbf{h}_k} \equiv A_{\ell k},
\end{equation}
so that $D = A$, and the matrix of brackets between the constraints $\Theta = (\bm{\phi}, \bm{\pi})$ and its inverse \eqref{constraint_matrix_blocks} are
\begin{equation}
\label{constraint_matrix_network}
\mathcal{M} = \begin{bmatrix} 0 & A \\ -A^T & 0 \end{bmatrix}, \qquad
\mathcal{M}^{-1} = \begin{bmatrix} 0 & -A^{-T} \\ A^{-1} & 0 \end{bmatrix}.
\end{equation}
Since $\bm{\phi}_\ell$ depends only on the activations of layers $\ell$ and $\ell - 1$, the matrix $A$ is block lower bidiagonal,
\begin{equation}
\label{a_matrix}
A = \begin{bmatrix}
\mathbb{I} & & & \\
-J_2 & \mathbb{I} & & \\
& \ddots & \ddots & \\
& & -J_L & \mathbb{I}
\end{bmatrix},
\end{equation}
with $\det A = 1$. Since the samples of the data set do not interact, each block is itself block diagonal over samples, and the constraint algebra splits into independent copies, one for each sample. The constraints \eqref{layer_constraints} are second class for any network architecture and any choice of parameters, and the Dirac bracket
\begin{equation}
\label{dirac_bracket_network}
\Delta(f, g) = \Gamma(f, g) - \Gamma(f, \Theta)\,\mathcal{M}^{-1}\,\Gamma(\Theta, g)
\end{equation}
is always well defined. Written in terms of the blocks of $\mathcal{M}^{-1}$, the time evolution generated by the Dirac bracket reads
\begin{equation}
\label{dirac_evolution_network}
\begin{split}
\frac{df}{dt} ={}& \Gamma(f, \mathcal{H}) + \Gamma(f, \bm{\phi})\,A^{-T}\,\Gamma(\bm{\pi}, \mathcal{H})\\
&- \Gamma(f, \bm{\pi})\,A^{-1}\,\Gamma(\bm{\phi}, \mathcal{H}) + \frac{\partial f}{\partial t},
\end{split}
\end{equation}
where we have assumed that the constraints do not depend explicitly on time; we return to this point at the end of the section.

\subsection{Backward and forward passes}
We now apply equation \eqref{dirac_evolution_network} to the parameters, their momenta and the activations. The brackets we need are $\Gamma(\bm{\pi}, \mathcal{H}) = -\partial c/\partial \mathbf{h}$, whose only nonvanishing block is the last one, and
\begin{equation}
\label{phi_h_bracket}
\Gamma(\bm{\phi}, \mathcal{H}) = \frac{\partial \bm{\phi}}{\partial \mathbf{x}} \frac{\partial \mathcal{H}}{\partial \mathbf{p}} = B\,\dot{\mathbf{x}}, \qquad B \equiv \frac{\partial \bm{\phi}}{\partial \mathbf{x}},
\end{equation}
where $B$ is block diagonal, with blocks $-G_\ell$.

\emph{Parameters.} The constraints \eqref{layer_constraints} contain no momenta $\mathbf{p}$, so $\Gamma(\mathbf{x}, \bm{\phi}) = \Gamma(\mathbf{x}, \bm{\pi}) = 0$, and the equation of motion of the parameters is unchanged by the constraints: $\Delta(\mathbf{x}, \mathcal{H}) = \Gamma(\mathbf{x}, \mathcal{H})$.

\emph{Momenta: backpropagation.} Since $\Gamma(\mathbf{p}, \bm{\phi}) = -B^T$ and $\Gamma(\mathbf{p}, \bm{\pi}) = 0$, equation \eqref{dirac_evolution_network} gives
\begin{equation}
\label{dirac_momenta_network}
\frac{d\mathbf{p}}{dt} = -\frac{\partial \mathcal{H}_0}{\partial \mathbf{x}} + B^T \bm{\delta}, \qquad \bm{\delta} \equiv A^{-T} \frac{\partial c}{\partial \mathbf{h}}.
\end{equation}
The vector $\bm{\delta} = (\bm{\delta}_1, \dots, \bm{\delta}_L)$ solves $A^T \bm{\delta} = \partial c/\partial \mathbf{h}$, and, since $A^T$ is block upper bidiagonal, it is obtained by back-substitution starting from the output layer:
\begin{equation}
\label{backprop_recursion}
\bm{\delta}_L = \frac{\partial c}{\partial \mathbf{h}_L}, \qquad \bm{\delta}_\ell = J_{\ell+1}^T\,\bm{\delta}_{\ell+1}, \quad \ell = L-1, \dots, 1.
\end{equation}
This is precisely the backward pass of the backpropagation algorithm, and $\bm{\delta}_\ell$ are the errors (adjoints) that it propagates through the layers, one for each sample. In the same way, the last term of equation \eqref{dirac_momenta_network} has the components $-G_\ell^T \bm{\delta}_\ell = -\partial L/\partial W_\ell$, in which the contributions of all samples add up; this is the gradient that backpropagation computes from the errors. Restoring the copy index, we find
\begin{equation}
\label{dirac_momenta_network_cov}
\begin{split}
\frac{dp^a}{dt} &= -\partial^a \mathcal{H}_0 - \big(\nabla L(\mathbf{x}_1), \nabla L(\mathbf{x}_2)\big)\\
&= -\partial^a\big(\mathcal{H}_0 + V\big), \qquad V = L(\mathbf{x}_1) - L(\mathbf{x}_2),
\end{split}
\end{equation}
together with $dx^a/dt = \bpartial^a \mathcal{H}_0$. These are Hamilton's equations \eqref{hamilton_equations_cov} for the Hamiltonian \eqref{hamiltonian_cov} with the loss \eqref{network_loss}, and in the physical limit they become equations \eqref{hamilton_equations_euc}. Equation \eqref{dirac_momenta_network_cov} holds exactly in the doubled phase space, not only in the physical limit. This follows from a general property of Dirac brackets: restricted to the constraint surface, the Dirac bracket is the Poisson bracket of the symplectic form induced on that surface. Here the constraint surface is parametrized by the doubled parameters and their momenta, since on it the activations of each copy are functions of the parameters of that copy, $\mathbf{h}_{\ell, i} = \mathbf{h}_\ell(\mathbf{x}_i)$, and their momenta vanish. The induced symplectic form is therefore the canonical one in $(\mathbf{x}_i, \mathbf{p}_i)$, and the Hamiltonian \eqref{network_hamiltonian} restricted to the surface is $\mathcal{H}_0 + L(\mathbf{x}_1) - L(\mathbf{x}_2)$. In other words, the Dirac bracket of the constrained network reproduces the full gradient of the loss, even though the Hamiltonian \eqref{network_hamiltonian} contains the loss only as a function of the network output: the backward pass is the inversion of the constraint matrix \eqref{constraint_matrix_network}.

\emph{Activations: forward-mode differentiation.} Since $\Gamma(\mathbf{h}, \bm{\phi}) = 0$ and $\Gamma(\mathbf{h}, \bm{\pi}) = \mathbb{I}$, equation \eqref{dirac_evolution_network} gives
\begin{equation}
\label{dirac_activations_network}
\frac{d\mathbf{h}}{dt} = -A^{-1} B\,\dot{\mathbf{x}}.
\end{equation}
Because $A$ is block lower bidiagonal, equation \eqref{dirac_activations_network} is solved by forward substitution,
\begin{equation}
\label{forward_mode_recursion}
\dot{\mathbf{h}}_\ell = J_\ell\,\dot{\mathbf{h}}_{\ell-1} + G_\ell\,\dot{W}_\ell, \qquad \dot{\mathbf{h}}_0 = 0,
\end{equation}
which is the Jacobian--vector product of forward-mode automatic differentiation. Moreover, since $A\,d\mathbf{h} + B\,d\mathbf{x} = 0$ on the constraint surface, equation \eqref{dirac_activations_network} states that $\dot{\mathbf{h}} = (\partial \mathbf{h}/\partial \mathbf{x})\,\dot{\mathbf{x}}$: the activations follow the parameters during training, so that the constraints remain satisfied along the motion.

The two off-diagonal blocks of $\mathcal{M}^{-1}$ thus generate the two modes of automatic differentiation: the block $A^{-1}$ propagates tangents forward through the network and drives the activations, while the block $A^{-T}$ propagates errors backward and drives the momenta of the parameters.

\emph{The Dirac bracket of the network.} The content of the constraint algebra can be stated compactly through the Dirac brackets of the activations. Since $\Gamma(\mathbf{h}, \Theta) = (0, \mathbb{I})$ and $\Gamma(\Theta, \mathbf{p}) = (B, 0)$, equation \eqref{dirac_bracket_network} gives
\begin{equation}
\label{network_dirac_jacobian}
\Delta(\mathbf{h}_\ell, \mathbf{p}) = -\big(A^{-1}B\big)_\ell = \frac{\partial \mathbf{h}_\ell}{\partial \mathbf{x}}, \qquad \Delta(\mathbf{h}_\ell, \mathbf{x}) = 0,
\end{equation}
and hence $\Delta\big(c(\mathbf{h}_L), \mathbf{p}\big) = \nabla L$. The Dirac bracket of an activation with the momenta of the parameters is the Jacobian of that activation with respect to the parameters: the constraint algebra encodes the full differential of the network, and the Dirac bracket is the chain rule written as a bracket. Forward-mode differentiation and backpropagation are its two contractions, with the velocities $\dot{\mathbf{x}}$ on the right, which gives the Jacobian--vector product \eqref{dirac_activations_network}, and with the gradient of the cost on the left, which gives the vector--Jacobian product \eqref{dirac_momenta_network}. The algebra itself does not prefer either mode; reverse mode is preferred in practice because the cost is a scalar. Physically, the fact that the constraints are second class means that they introduce no gauge freedom: the activations are completely determined by the parameters, the reduced phase space is that of the parameters alone, and the Dirac bracket is the bracket induced on it, as when the coordinates of the rigidly connected parts of a mechanism are eliminated.

\emph{Lagrange multipliers.} Equivalently, the Dirac evolution \eqref{dirac_evolution_network} is generated by the total Hamiltonian $\mathcal{H}_T = \mathcal{H} + \mathbf{u}_{\phi}\cdot\bm{\phi} + \mathbf{u}_{\pi}\cdot\bm{\pi}$, with the Lagrange multipliers fixed by the requirement that the constraints be preserved in time. Comparing with equations \eqref{dirac_momenta_network} and \eqref{dirac_activations_network}, the multipliers are
\begin{equation}
\label{network_multipliers}
\mathbf{u}_{\phi} = -\bm{\delta}, \qquad \mathbf{u}_{\pi} = \dot{\mathbf{h}}:
\end{equation}
the multipliers of the layer constraints are, up to sign, the errors of backpropagation, and the multipliers of the momentum constraints are the velocities of the activations.

\subsection{Mechanical interpretation}
The kinematic chain introduced above makes these results intuitive. Two classical problems are associated with any mechanism. The first is velocity kinematics: given the velocities of the joints, find the velocity of every link. Each link moves because the link before it moves and because its own joint moves, which is the forward recursion \eqref{forward_mode_recursion}, with $J_\ell$ the transmission ratio between consecutive links and $G_\ell$ the effect of joint $\ell$. The second is statics: given a load on the end effector, find the forces carried by the links and the generalized forces felt by the joints. The load is the force $-\partial c/\partial\mathbf{h}_L$ of the potential energy $c$; it is transmitted back toward the base, link by link, by the transposed transmission ratios, and the joints feel the generalized forces $-G_\ell^T\bm{\delta}_\ell$. This is the backward recursion \eqref{backprop_recursion}: up to sign, the errors $\bm{\delta}_\ell$ are the forces carried by the links, and the gradient of the loss is the generalized force exerted by the load on the joints.

The two problems are related by the principle of virtual work. The power delivered by the load, $-(\partial c/\partial\mathbf{h}_L)\cdot\dot{\mathbf{h}}_L$, must equal the power delivered to the joints, so the map from loads to joint forces is the transpose of the map from joint velocities to link velocities. In robotics this is the familiar statement that the joint torques required to balance a force on the end effector are obtained by applying the transpose of the Jacobian of the arm to that force. Here it is the statement that the two off-diagonal blocks of the inverse constraint matrix \eqref{constraint_matrix_network} are $A^{-1}$ and $A^{-T}$. Finally, constraint forces in mechanics are Lagrange multipliers, and equation \eqref{network_multipliers} identifies them with the errors. From this point of view, backpropagation is the computation of the constraint forces of a rigid mechanism, and training is the damped motion of its joints under the load of the cost.

\subsection{Remarks}
\emph{Independence of the optimizer.} The constraints \eqref{layer_constraints} involve only the parameters, the activations and the momenta of the activations; they contain neither the momenta $\mathbf{p}$ nor $\mu$, $\gamma$ or $\mathcal{H}_0$. The constraint matrix \eqref{constraint_matrix_network}, its inverse and the Dirac bracket \eqref{dirac_bracket_network} are therefore properties of the network, that is, of its architecture, its parameters and its inputs, and not of the optimizer. Their functional form is the same for every optimizer, although their values depend on the point of the trajectory at which they are evaluated, and different optimizers follow different trajectories. The optimizer enters only through the velocities $\dot{\mathbf{x}}$ on which the forward mode acts, through $\Gamma(\bm{\phi}, \mathcal{H}) = B\dot{\mathbf{x}}$, and through the remaining force $-\partial\mathcal{H}_0/\partial\mathbf{x}$ on the momenta; the errors $\bm{\delta} = A^{-T}\partial c/\partial\mathbf{h}$ do not depend on it at all. The derivation above therefore applies unchanged to every optimizer described by the Hamiltonian \eqref{hamiltonian_cov}, with arbitrary time-dependent mass $\mu(t)$ and damping coefficient $\gamma(t)$. Two optimizers enlarge the constraint algebra with constraints of their own, while leaving the block of the network unchanged. The first is Nesterov's method. Its Hessian-damped Hamiltonian contains the gradient of the loss explicitly, but the look-ahead Hamiltonian \eqref{nesterov_lookahead_hamiltonian} depends on the loss only through its values at the look-ahead points $\mathbf{y}_i = \mathbf{x}_i + h\bar{\mathbf{v}}$. The construction above then applies with one extra layer in front of the network: the look-ahead parameters $\mathbf{y}_i$ become coordinates, with vanishing momenta and the constraints $\bm{\psi}_i \equiv \mathbf{y}_i - \mathbf{x}_i - h\bar{\mathbf{v}} \approx 0$, and the layer constraints \eqref{layer_constraints} are imposed with the parameters $\mathbf{y}_i$ in place of $\mathbf{x}_i$. Unlike the layer constraints, the look-ahead constraints depend on the momenta, through $\bar{\mathbf{v}}$, so that the Dirac bracket of the parameters with the Hamiltonian is no longer equal to the Poisson bracket. Nevertheless, the Dirac evolution of $\mathbf{x}_i$ and $\mathbf{p}_i$ coincides, in the whole doubled space, with Hamilton's equations for the Hamiltonian \eqref{nesterov_lookahead_hamiltonian} with the loss \eqref{network_loss}. The reason is general: the Dirac bracket restricted to the constraint surface is the Poisson bracket of the symplectic form induced on that surface; here the surface is parametrized by $(\mathbf{x}_i, \mathbf{p}_i)$, all the constrained momenta vanish on it, and the induced form is the canonical one, while the Hamiltonian \eqref{network_hamiltonian} restricted to the surface is $\ham_{\text{LA}}$. The backward pass is then evaluated at the look-ahead point, as prescribed by Nesterov's method. Unlike the layer constraints, the look-ahead constraints of the two copies do not commute, $\Gamma(\bm{\psi}_1, \bm{\psi}_2) = (h/\mu)\mathbb{I}$, so that the two copies can no longer be treated separately, and the activations of the two copies no longer commute with each other. These results are derived in Appendix~\ref{appendix_nesterov_lookahead}. The second is gradient flow, whose parameters carry, in addition, the constraints \eqref{second_class_constraints_gradient_flow}, which have nonvanishing brackets with $\bm{\phi}$. Dirac brackets can, however, be constructed in stages~\cite{henneaux1992quantization}: imposing the constraints \eqref{layer_constraints} first replaces the cost of the network output by the loss \eqref{network_loss}, and imposing the constraints \eqref{second_class_constraints_gradient_flow} afterwards reproduces the analysis of Appendix~\ref{appendix_hamiltonian_mechanics_gradient_flow}.

\emph{Mini-batches.} When the inputs $\mathbf{h}_0^{(n)}$ and targets $\mathbf{y}^{(n)}$ change from one mini-batch to the next, the constraints $\bm{\phi}_1$ and the Hamiltonian acquire an explicit time dependence. The evolution then contains the last term of equation \eqref{dirac_evolution_time_dependent}, which reduces to $-\Gamma(f, \bm{\pi})\,A^{-1}\,\partial\bm{\phi}/\partial t$. It changes the evolution of the activations, but not that of the parameters and their momenta, which have vanishing brackets with $\bm{\pi}$.

\emph{Scope of the result.} Two ingredients of the construction should be distinguished. The first is general: for any Hamiltonian of the form $\mathcal{H}_0 + c(\mathbf{h}_L)$ and any constraints that determine the activations as functions of the parameters, the Dirac bracket restricted to the constraint surface is the bracket of the reduced phase space, and the reduced Hamiltonian contains the loss \eqref{network_loss}. That the Dirac evolution produces the gradient of the loss is therefore guaranteed; what the construction adds is the explicit form of the inverse constraint matrix \eqref{constraint_matrix_network}, in which the backward recursion \eqref{backprop_recursion}, the forward-mode recursion \eqref{forward_mode_recursion} and the multipliers \eqref{network_multipliers} appear as the solution of a triangular system. The second ingredient is specific to training: optimizers are dissipative, and without the Bateman--Galley doubling they have no Hamiltonian in which the layers could be imposed as constraints. The doubling supplies it, and the results of this section then hold for every optimizer described by the Hamiltonian \eqref{hamiltonian_cov} with arbitrary $\mu(t)$ and $\gamma(t)$, in the whole doubled phase space and not only in the physical limit, for Nesterov's method through the momentum-dependent look-ahead constraints, and for gradient flow through the Dirac bracket in stages.

\emph{Relation to previous work.} The backward recursion \eqref{backprop_recursion} coincides with the equations satisfied by the Lagrange multipliers in the Lagrangian formulation of backpropagation~\cite{lecun1988theoretical} and in the adjoint method~\cite{chen2018neural}. In those formulations, the multipliers enforce the layer equations along the depth of the network, whereas here the layer equations are constraints on the training dynamics, and the errors appear as matrix elements of the inverse constraint matrix. Backpropagation has also been derived from a Bateman--Galley action in which the depth of the network plays the role of time, and in which the doubled variables encode the activations and the errors~\cite{scurria2026physical}. In contrast, in our formulation the time is the training time, the network is described by holonomic constraints, and backpropagation follows from the Dirac bracket of the optimizer itself.

\emph{Worked example.} Appendix~\ref{appendix_two_layer_example} works out the construction for a dense network with one hidden layer, the simplest case in which the backward pass is nontrivial. For the linear model of least-squares regression, $L = 1$, the matrix $A$ is the identity and there is no backward recursion.

\section{Numerical test}
\label{section_numerical_test}
\label{section_numerical_experiments}
\begin{figure*}[tb]
\includegraphics[width=0.8\textwidth]{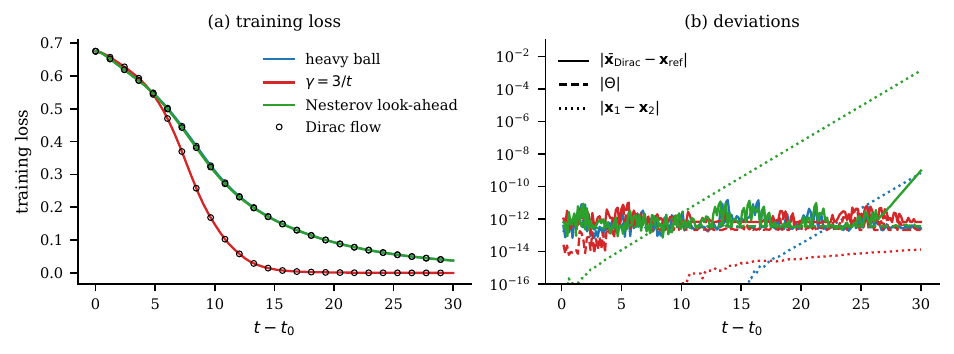}
\caption{Training a $3$--$5$--$4$--$2$ network by integrating the Dirac flow of the constrained doubled system, whose Hamiltonian contains only the cost of the network output. (a) Training loss from the Dirac flow (circles) and from ordinary training with reverse-mode gradients (lines), for the heavy-ball method, $\gamma = 3/t$ and Nesterov's look-ahead dynamics with $h = 0.1$, whose curve nearly coincides with that of the heavy-ball method. (b) Deviation of the copy-averaged parameters from the reference (solid), constraint violation (dashed), and separation of the two copies (dotted), which grows from round-off through the anti-damped mirror modes.}
\label{fig_dirac_backprop}
\end{figure*}
We test the construction of Sec.~\ref{section_backpropagation} numerically, to high precision, on a small network. The code is archived on Zenodo~\cite{holanda2026dirac} and developed at \url{https://github.com/linneuholanda/dirac-backprop-code}.

The data are synthetic. The inputs are $N = 6$ vectors $\mathbf{h}_0 \in \mathbb{R}^3$ with independent standard normal entries. For the squared-error cost, the targets in $\mathbb{R}^2$ are also standard normal; for the cross-entropy cost, they are one-hot labels of two classes, chosen uniformly at random. The weights are initialized with independent normal entries of variance $1/n_{\text{in}}$, where $n_{\text{in}}$ is the number of inputs of the layer, and the trajectories start at rest. All trajectories are integrated with an adaptive eighth-order Runge--Kutta method at relative tolerance $10^{-11}$, in double precision. The pointwise tests below are performed at points $\mathbf{x}_i$ obtained by adding $0.3$ times standard normal noise to the initial weights, with momenta $\mathbf{p}_i$ with standard normal entries, independently for the two copies $i = 1, 2$, at $t = 2, 3, 4$. We repeated every test with four further random seeds, and all checks passed again with the same tolerances.

To test the construction of Sec.~\ref{section_backpropagation} independently of its derivation, we implement the Dirac bracket generically. A point of the enlarged doubled phase space collects the parameters, the activations of all layers and samples, and their momenta, for both copies. The bracket $\Gamma = \Gamma_1 - \Gamma_2$ is a constant matrix $\Omega$, the gradients of the constraints \eqref{layer_constraints} are obtained by automatic differentiation of the layer equations alone, and the evolution is
\begin{equation}
\label{numerical_dirac_evolution}
\begin{gathered}
\dot{z} = \Omega\,\partial\mathcal{H} - \Omega\,(\partial\Theta)^T\,\mathcal{M}^{-1}\Big((\partial\Theta)\,\Omega\,\partial\mathcal{H} + \frac{\partial\Theta}{\partial t}\Big),\\
\mathcal{M} = (\partial\Theta)\,\Omega\,(\partial\Theta)^T,
\end{gathered}
\end{equation}
with the Hamiltonian \eqref{network_hamiltonian}, which contains the cost of the network output only. No backward pass appears anywhere in this computation. The reference is ordinary training: Hamilton's equations of $\mathcal{H}_0 + L(\mathbf{x}_1) - L(\mathbf{x}_2)$, or of $\mathcal{H}_{\text{LA}}$, with the gradient of the loss computed by reverse-mode automatic differentiation of the whole network. The network has two hidden layers with hyperbolic-tangent activations and a linear output layer, with widths $3$--$5$--$4$--$2$ and no biases, so that $43$ parameters and, for $N = 6$ samples, $66$ activations per copy span a constrained phase space of dimension $436$ with $264$ second-class constraints. With the look-ahead layer of Nesterov's method the dimension is $608$, with $436$ constraints.

At random points of the constraint surface away from the physical limit ($\mathbf{x}_1 \neq \mathbf{x}_2$, $\mathbf{p}_1 \neq \mathbf{p}_2$), for the squared-error and cross-entropy costs and for the heavy-ball method and $\gamma = 3/t$, we find agreement to within $2\times 10^{-15}$ for every statement of Sec.~\ref{section_backpropagation}: the evolution of the parameters and momenta \eqref{dirac_momenta_network_cov}, the evolution of the activations \eqref{dirac_activations_network}, which equals the forward-mode Jacobian--vector product, the Dirac brackets \eqref{network_dirac_jacobian} of the activations with the momenta, which equal $+\partial\mathbf{h}/\partial\mathbf{x}$ for the first copy and $-\partial\mathbf{h}/\partial\mathbf{x}$ for the second, the vanishing brackets with the parameters, and the Lagrange multipliers \eqref{network_multipliers}, which equal $-\bm{\delta}$ for the first copy and $+\bm{\delta}$ for the second, with $\bm{\delta}$ the errors computed by reverse-mode automatic differentiation. The same agreement holds for Nesterov's look-ahead layer, whose Dirac evolution reproduces Hamilton's equations of $\mathcal{H}_{\text{LA}}$ \eqref{nesterov_lookahead_hamiltonian}; for $\gamma = 3/t$ its constraints depend explicitly on time, but the term $\partial\Theta/\partial t$ of equation \eqref{numerical_dirac_evolution} affects only the evolution of the look-ahead coordinates, not that of $\mathbf{x}$ and $\mathbf{p}$. As a negative control, the Poisson bracket alone, without the constraints, misses the gradient of the loss entirely. The statements of Appendix~\ref{appendix_nesterov_lookahead}, which are specific to the look-ahead constraints, hold to within $3\times 10^{-15}$ at the same points: the Dirac brackets \eqref{lookahead_dirac_brackets}, including $\Delta(\mathbf{y}_1, \mathbf{y}_2) = (h/\mu)\mathbb{I}$ between the look-ahead points of the two copies; the multipliers \eqref{lookahead_multipliers} and \eqref{lookahead_multipliers_momenta}, that is, the errors and gradients at the look-ahead point; the evolution \eqref{lookahead_constrained_evolution} of the look-ahead points and activations; and the terms proportional to $\mathbf{g}_-$ in equations \eqref{lookahead_equations_of_motion}, which vanish in the physical limit but not at these points, where the largest entry of $\mathbf{g}_-$ is at least $0.16$.

Finally, we integrate the Dirac flow \eqref{numerical_dirac_evolution} in the full constrained phase space, starting on the physical limit, for the cross-entropy cost, and compare it with the ordinary integration of equation \eqref{ode} and, for Nesterov's method, of equation \eqref{nesterov_lookahead_ode} with $h = 0.1$. The two descriptions produce the same training (Fig.~\ref{fig_dirac_backprop}). The constraints remain satisfied to $1.4\times 10^{-11}$, and while the two copies are close, the parameters, averaged over the copies, agree with the reference to $1.5\times 10^{-11}$. Round-off errors excite the anti-damped mirror modes of the doubled system, and the two copies separate exponentially, here at the rate $1.0 = \gamma/\mu$ of a mirror mode along a nearly flat direction of the loss (for $\gamma = 3/t$ the damping decays, and the separation stays below $10^{-13}$). The separation affects the copy average only at second order: for Nesterov's method, the copies are $1.4\times 10^{-3}$ apart at $t = 30$, and their average deviates from the reference by $10^{-9}$.

\section{Conclusion}
\label{section_conclusion}

We have shown that, once the Bateman--Galley doubling gives gradient-based optimizers a Hamiltonian, backpropagation is Dirac evolution. Imposing the layers of a feed-forward network as constraints \eqref{layer_constraints}, the Hamiltonian \eqref{network_hamiltonian} needs to contain only the cost of the network output. The constraints are second class for every architecture, the Dirac bracket of an activation with the momenta of the parameters is its Jacobian \eqref{network_dirac_jacobian}, the backward pass \eqref{backprop_recursion} is the back-substitution that inverts the constraint matrix, and the errors are the Lagrange multipliers \eqref{network_multipliers} of the layer constraints, that is, the constraint forces of a kinematic chain under the load of the cost. These statements hold exactly in the doubled phase space. The network's part of the constraint algebra is shared by all the optimizers considered here. Nesterov's look-ahead enters as one additional layer of momentum-dependent constraints, which couple the two copies of the doubled system, although the effect of this coupling on the equations of motion vanishes in the physical limit. Gradient flow is obtained by constructing the Dirac bracket in stages. The numerical test of Sec.~\ref{section_numerical_test} confirms these statements pointwise to machine precision and trains a network by integrating the Dirac flow, in agreement with ordinary training to $10^{-11}$, with no backward pass in the computation.

The formalism describes the continuum limit of the optimizers, deterministic training on the full data set, and feed-forward networks; mini-batches enter only as an explicit time dependence of the constraints. The physical limit is unstable, because the mirror copy is anti-damped, so that the doubled system is a device for deriving the equations of motion and brackets, and numerical integrations must take the separation of the copies into account. The construction relies only on the fact that each activation is determined by earlier activations and by the parameters, which suggests that it extends to residual connections, weight sharing and attention layers, and more generally to any computational graph without cycles. Since the Bateman--Galley doubling is the classical limit of the closed-time-path formalism~\cite{galley2013classical}, which also describes fluctuations, it is also a natural starting point for a treatment of stochastic training. Finally, an action in which both the training time and the depth of the network play the role of time could connect the present formulation with that of Ref.~\cite{scurria2026physical}.

\section*{Data availability}
The code that produces Fig.~\ref{fig_dirac_backprop} and the numbers in Sec.~\ref{section_numerical_test} is openly available on Zenodo~\cite{holanda2026dirac}. The data are synthetic and are generated by this code.

\begin{acknowledgments}
Portions of this work were carried out with the assistance of Claude Opus 5.5 (Anthropic), an AI model, used through Claude Code. Under the authors' direction, it helped draft and edit the manuscript. It also wrote the symbolic (SymPy) checks of the derivations and the code for the numerical test, which produces Fig.~\ref{fig_dirac_backprop} and the numbers in Sec.~\ref{section_numerical_test}. The authors directed this work, checked the derivations and results, and take full responsibility for the content of the paper.
\end{acknowledgments}

\appendix

\section{Hamiltonian mechanics of gradient flow}
\label{appendix_hamiltonian_mechanics_gradient_flow}

In this appendix, we derive the Hamiltonian dynamics of gradient flow within the Bateman--Galley formalism. The Lagrangian that gives rise to equation \eqref{gradient_flow} follows immediately from equations \eqref{lagrangian} and \eqref{lagrangian_cov} by setting the mass $\mu(t)$ to zero:
\begin{align}
\label{gradient_flow_lagrangian_euclid}
\lagr =& -L(\mathbf{x}_1) + L(\mathbf{x}_2) + \frac{1}{4}\dot{\gamma}(t)(\mathbf{x}_1^2 - \mathbf{x}_2^2) \nonumber\\
&+ \frac{1}{2}\gamma(t)(\dot{\mathbf{x}}_1\cdot\mathbf{x}_2 - \dot{\mathbf{x}}_2\cdot\mathbf{x}_1),
\end{align}
which in Minkowski notation becomes
\begin{equation}
\label{gradient_flow_lagrangian_minkowski}
\lagr(x, \dot{x}, t) = -V + \frac{1}{4}\dot{\gamma}(t)x^ax_a + \frac{1}{2}\gamma(t)\dot{x}^ax^{\dagger}_a.
\end{equation}
Applying the Euler--Lagrange equations \eqref{euler_lagrange_eq_cov} to the Lagrangian \eqref{gradient_flow_lagrangian_minkowski}, we obtain the equations of motion of the doubled system,
\begin{equation}
\label{gradient_flow_euler_lagrange_doubled}
\gamma(t)\,\dot{x}^{\dagger a} = -\partial^a V + \frac{\dot{\gamma}(t)}{2}\big(x^a - {x^{\dagger}}^a\big),
\end{equation}
or, in Euclidean form,
\begin{subequations}
\begin{align}
\gamma(t)\,\dot{\mathbf{x}}_2 &= -\nabla L(\mathbf{x}_1) + \frac{\dot{\gamma}(t)}{2}(\mathbf{x}_1 - \mathbf{x}_2),\\
\gamma(t)\,\dot{\mathbf{x}}_1 &= -\nabla L(\mathbf{x}_2) - \frac{\dot{\gamma}(t)}{2}(\mathbf{x}_1 - \mathbf{x}_2).
\end{align}
\end{subequations}
As in the massive case, the two copies are coupled: the velocity of each copy is driven by the gradient of the loss evaluated at the other. Multiplying equation \eqref{gradient_flow_euler_lagrange_doubled} by $X$ and using $X\dot{x}^{\dagger} = \dot{x}$, we can solve for the generalized velocities,
\begin{equation}
\label{gradient_flow_velocities_doubled}
\dot{x}^a = -\frac{1}{\gamma(t)}X^a{}_b\,\partial^bV + \frac{\dot{\gamma}(t)}{2\gamma(t)}\big({x^{\dagger}}^a - x^a\big).
\end{equation}
In the physical limit, ${x^{\dagger}}^a - x^a$ vanishes and $X^a{}_b\,\partial^bV$ reduces to $\nabla L$ in both components, so that equation \eqref{gradient_flow_velocities_doubled} yields the physical (undoubled) equation of motion
\begin{equation}
\label{gradient_flow_lagrangian_physical_limit}
\frac{d\mathbf{x}}{dt} = -\frac{1}{\gamma(t)}\nabla L,
\end{equation}
which is gradient flow, equation \eqref{gradient_flow}, with a damping coefficient that may depend on time.

The Hamiltonian of gradient flow then follows from the Legendre transformation \eqref{legendre_transformation_cov}:
\begin{equation}
\label{gradient_flow_hamiltonian_minkowski}
\mathcal{H} = V - \frac{1}{4}\dot{\gamma}(t)x^ax_a.
\end{equation}
Note that the system described by the Lagrangian \eqref{gradient_flow_lagrangian_minkowski} (or, equivalently, by the Hamiltonian \eqref{gradient_flow_hamiltonian_minkowski}) is more general than standard gradient flow, since the damping coefficient $\gamma(t)$, and hence the learning rate, may depend on time. It is also clear from equation \eqref{gradient_flow_hamiltonian_minkowski} that the resulting Hamiltonian has no kinetic term, which is a hallmark of constrained systems. This can also be seen from the fact that the conjugate momenta of gradient flow,
\begin{equation}
    \label{conjugate_momenta_gradient_flow}
    p^a = \frac{1}{2}\gamma(t){x^{\dagger}}^a,
\end{equation}
do not depend on the generalized velocities $\dot{x}^a$.

In general, systems whose conjugate momenta define second-class constraints in phase space can be consistently described by a Hamiltonian formalism only through Dirac's generalized theory of constrained Hamiltonian systems~\cite{dirac1950generalized}. Heavily damped systems whose dynamics are governed by gradient flow are an example: they are described by the Lagrangian \eqref{gradient_flow_lagrangian_euclid}, which is linear in the generalized velocities. In such cases, the map between canonical coordinates and generalized coordinates and velocities is not a bijection, and the Hamiltonian formalism for unconstrained systems based on Poisson brackets breaks down. In what follows, we show how to generalize the Poisson brackets \eqref{scalar_poisson_bracket} to the Dirac brackets needed to treat gradient flow dynamics.

Let $\Gamma(f,g)$ and $\Delta(f,g)$ be the Poisson and Dirac brackets between two scalars $f$ and $g$, respectively. Then
\begin{equation}
\label{dirac_brackets}
\Delta(f,g) = \Gamma(f,g) + \Gamma(f, \chi_a) \Gamma^{-1}(\chi^a,\chi_b) \Gamma(g,\chi^b).
\end{equation}
From equation \eqref{conjugate_momenta_gradient_flow}, the second-class constraints can be written as
\begin{equation}
\label{second_class_constraints_gradient_flow}
    \chi^a = \frac{\gamma(t)}{2}{x^{\dagger}}^a - p^a,
\end{equation}
so that the Dirac bracket of gradient flow takes the general form
\begin{equation}
\label{dirac_brackets_gradient_flow}
\Delta(f,g) = \Gamma(f,g) -\frac{1}{\gamma(t)} \Gamma(f, \chi_a) X^a{}_b \Gamma(g,\chi^b),
\end{equation}
where we have used the Pauli $X$ matrix in doubled dimensions, equation \eqref{cal_x}.

Since the constraints \eqref{second_class_constraints_gradient_flow} depend explicitly on time through $\gamma(t)$, the Dirac bracket alone does not preserve them along the motion. The time evolution of a phase-space function $f$ then acquires an additional term~\cite{gitman1990quantization},
\begin{equation}
\label{dirac_time_evolution_gradient_flow}
\frac{df}{dt} = \Delta(f,\mathcal{H}) + \frac{\partial f}{\partial t} + \frac{1}{\gamma(t)}\Gamma(f,\chi_a)X^a{}_b\frac{\partial\chi^b}{\partial t},
\end{equation}
which ensures that $\dot{\chi}^a = 0$ holds identically. It is the general relation \eqref{dirac_evolution_time_dependent} of Sec.~\ref{section_constraint_algebra}, applied to the constraints \eqref{second_class_constraints_gradient_flow}. For a time-independent damping coefficient, $\partial\chi^b/\partial t = 0$ and equation \eqref{dirac_time_evolution_gradient_flow} reduces to the familiar form $\dot{f} = \Delta(f,\mathcal{H}) + \partial f/\partial t$. From equations \eqref{dirac_brackets_gradient_flow} and \eqref{dirac_time_evolution_gradient_flow}, it is straightforward to show that Hamilton's equations of motion of gradient flow are
\begin{subequations}
\label{hamiltons_equations_of_motion_gradient_flow}
    \begin{equation}
        \dot{x}^a = -\frac{1}{\gamma(t)}X^a{}_b\,\partial^bV + \frac{\dot{\gamma}(t)}{2\gamma(t)}\big({x^{\dagger}}^a - x^a\big),
    \end{equation}
    \begin{equation}
        \dot{p}^a = - \frac{1}{2}\partial^a V + \frac{\dot{\gamma}(t)}{4}\big(x^a + {x^{\dagger}}^a\big),
    \end{equation}
\end{subequations}
which become, in the physical limit,
\begin{subequations}
\label{equations_of_motion_gradient_flow_physical_limit}
    \begin{equation}
    \frac{d\mathbf{x}}{dt} = -\frac{1}{\gamma(t)}\nabla L,
    \end{equation}
    \begin{equation}
    \frac{d\mathbf{p}}{dt} = -\frac{1}{2}\nabla L + \frac{\dot{\gamma}(t)}{2}\mathbf{x}.
    \end{equation}
\end{subequations}
Equation (\ref{hamiltons_equations_of_motion_gradient_flow}a) coincides with equation \eqref{gradient_flow_velocities_doubled}, obtained directly from the doubled Euler--Lagrange equations, so the Hamiltonian and Lagrangian descriptions of gradient flow agree in the doubled space, not only in the physical limit. Accordingly, equation (\ref{equations_of_motion_gradient_flow_physical_limit}a) recovers equation \eqref{gradient_flow_lagrangian_physical_limit}, gradient flow with the time-dependent learning rate $1/\gamma(t)$, and equation (\ref{equations_of_motion_gradient_flow_physical_limit}b) is simply the time derivative of the constraint $\mathbf{p} = \frac{1}{2}\gamma(t)\mathbf{x}$, the physical limit of equation \eqref{second_class_constraints_gradient_flow}. The standard form of gradient flow is the special case of a time-independent damping coefficient $\gamma$.

\section{Worked example: a two-layer dense network}
\label{appendix_two_layer_example}

In this appendix we work out the construction of Sec.~\ref{section_backpropagation} explicitly for the simplest nonlinear case: a dense network with one hidden layer and a linear output layer. It adds a hidden layer to the linear model of least-squares regression, and with it a nontrivial backward pass. We keep the cost generic, and at the end specialize it to least-squares regression and to classification with the cross-entropy. For clarity we consider a single sample; for a batch, each sample carries its own copy of the activations and constraints, and the gradients add up, as in Sec.~\ref{section_backpropagation}. As in Sec.~\ref{section_backpropagation}, the two Bateman--Galley copies obey identical equations, so we write all expressions for a single copy and use $\Gamma = \Gamma_1$. The latin indices $i, j, k$ in this appendix label components of vectors and matrices, not copies.

\subsection{Network, phase space and constraints}
Given an input $\mathbf{h}_0 \in \mathbb{R}^{n_0}$, the network computes
\begin{equation}
\label{two_layer_network}
\mathbf{z}_1 = W_1\mathbf{h}_0, \qquad \mathbf{h}_1 = \sigma(\mathbf{z}_1), \qquad \mathbf{h}_2 = W_2\mathbf{h}_1,
\end{equation}
where $W_1$ is an $n_1 \times n_0$ matrix, $W_2$ is an $n_2 \times n_1$ matrix, and the activation function $\sigma$ acts componentwise. The output $\mathbf{h}_2 \in \mathbb{R}^{n_2}$ is scored by a differentiable cost $c(\mathbf{h}_2)$, which also depends on the target associated with $\mathbf{h}_0$, so that the loss \eqref{network_loss} is
\begin{equation}
\label{two_layer_loss}
L(W_1, W_2) = c\big(W_2\,\sigma(W_1\mathbf{h}_0)\big).
\end{equation}
The layer Jacobians \eqref{layer_jacobians} are
\begin{equation}
\label{two_layer_jacobians}
\begin{gathered}
J_2 = W_2, \qquad \frac{\partial (\mathbf{h}_1)_k}{\partial (W_1)_{ij}} = \sigma'\big((\mathbf{z}_1)_k\big)\,(\mathbf{h}_0)_j\,\delta_{ki},\\
\frac{\partial (\mathbf{h}_2)_k}{\partial (W_2)_{ij}} = (\mathbf{h}_1)_j\,\delta_{ki},
\end{gathered}
\end{equation}
where $\delta_{ki}$ is the Kronecker delta.

The coordinates of phase space are the entries of $W_1$ and $W_2$, with conjugate momenta $P_1$ and $P_2$ (matrices of the same shapes), and the activations $\mathbf{h}_1$ and $\mathbf{h}_2$, with conjugate momenta $\bm{\pi}_1$ and $\bm{\pi}_2$. Poisson brackets sum over all entries; for example, $\Gamma\big((W_\ell)_{ij}, (P_m)_{kl}\big) = \delta_{\ell m}\,\delta_{ik}\,\delta_{jl}$ and $\Gamma\big((\mathbf{h}_\ell)_i, (\bm{\pi}_m)_k\big) = \delta_{\ell m}\,\delta_{ik}$. The Hamiltonian \eqref{network_hamiltonian} is
\begin{equation}
\label{two_layer_hamiltonian}
\mathcal{H} = \mathcal{H}_0(W, P, t) + c(\mathbf{h}_2),
\end{equation}
where $\mathcal{H}_0$ is the optimizer Hamiltonian of Sec.~\ref{section_hamiltonian_learning} for the entries of $W_1$ and $W_2$, without the potential. The constraints \eqref{layer_constraints} are
\begin{equation}
\label{two_layer_constraints}
\begin{gathered}
\bm{\phi}_1 = \mathbf{h}_1 - \sigma(W_1\mathbf{h}_0) \approx 0, \qquad \bm{\phi}_2 = \mathbf{h}_2 - W_2\mathbf{h}_1 \approx 0,\\
\bm{\pi}_1 \approx 0, \qquad \bm{\pi}_2 \approx 0 .
\end{gathered}
\end{equation}

\subsection{Constraint matrix}
The nonvanishing brackets between the constraints are
\begin{equation}
\label{two_layer_constraint_brackets}
\begin{gathered}
\Gamma(\bm{\phi}_1, \bm{\pi}_1) = \mathbb{I}, \qquad \Gamma(\bm{\phi}_2, \bm{\pi}_2) = \mathbb{I},\\
\Gamma(\bm{\phi}_2, \bm{\pi}_1) = -W_2,
\end{gathered}
\end{equation}
together with their antisymmetric counterparts, while $\Gamma(\bm{\phi}_1, \bm{\pi}_2) = 0$ because $\bm{\phi}_1$ does not depend on $\mathbf{h}_2$. Hence
\begin{equation}
\label{two_layer_a_matrix}
\begin{gathered}
A = \begin{bmatrix} \mathbb{I} & 0 \\ -W_2 & \mathbb{I} \end{bmatrix}, \qquad
A^{-1} = \begin{bmatrix} \mathbb{I} & 0 \\ W_2 & \mathbb{I} \end{bmatrix},\\
A^{-T} = \begin{bmatrix} \mathbb{I} & W_2^T \\ 0 & \mathbb{I} \end{bmatrix},
\end{gathered}
\end{equation}
and, ordering the constraints as $(\bm{\phi}_1, \bm{\phi}_2, \bm{\pi}_1, \bm{\pi}_2)$, the constraint matrix \eqref{constraint_matrix_network} and its inverse are
\begin{equation}
\label{two_layer_constraint_matrix}
\begin{gathered}
\mathcal{M} = \begin{bmatrix}
0 & 0 & \mathbb{I} & 0 \\
0 & 0 & -W_2 & \mathbb{I} \\
-\mathbb{I} & W_2^T & 0 & 0 \\
0 & -\mathbb{I} & 0 & 0
\end{bmatrix},\\
\mathcal{M}^{-1} = \begin{bmatrix}
0 & 0 & -\mathbb{I} & -W_2^T \\
0 & 0 & 0 & -\mathbb{I} \\
\mathbb{I} & 0 & 0 & 0 \\
W_2 & \mathbb{I} & 0 & 0
\end{bmatrix}.
\end{gathered}
\end{equation}
Note that $\mathcal{M}^{-1}$ involves only $W_2$, and no derivative of the activation function: the nonlinearity of the hidden layer enters the Dirac evolution only through the brackets of the constraints with the momenta $P_1$.

\subsection{Backward pass}
The cost depends only on $\mathbf{h}_2$, so $\Gamma(\bm{\pi}_1, \mathcal{H}) = 0$ and $\Gamma(\bm{\pi}_2, \mathcal{H}) = -\partial c/\partial\mathbf{h}_2$. The errors $\bm{\delta} = A^{-T}\,\partial c/\partial\mathbf{h}$ of equation \eqref{dirac_momenta_network} are then
\begin{equation}
\label{two_layer_errors}
\bm{\delta}_2 = \frac{\partial c}{\partial \mathbf{h}_2}, \qquad \bm{\delta}_1 = W_2^T\,\bm{\delta}_2,
\end{equation}
which is the backward recursion \eqref{backprop_recursion}: the output error is propagated to the hidden layer by the transpose of the output weights. The brackets of the momenta with the layer constraints follow from equation \eqref{two_layer_jacobians},
\begin{equation}
\label{two_layer_momentum_brackets}
\begin{gathered}
\Gamma\big((P_1)_{ij}, (\bm{\phi}_1)_k\big) = \sigma'\big((\mathbf{z}_1)_k\big)\,(\mathbf{h}_0)_j\,\delta_{ki},\\
\Gamma\big((P_2)_{ij}, (\bm{\phi}_2)_k\big) = (\mathbf{h}_1)_j\,\delta_{ki},
\end{gathered}
\end{equation}
and the remaining brackets of $P_1$, $P_2$ with the constraints vanish. Inserting equations \eqref{two_layer_errors} and \eqref{two_layer_momentum_brackets} into the Dirac evolution \eqref{dirac_evolution_network}, we find
\begin{subequations}
\label{two_layer_momenta_evolution}
\begin{equation}
\frac{dP_2}{dt} = -\frac{\partial \mathcal{H}_0}{\partial W_2} - \bm{\delta}_2\,\mathbf{h}_1^T,
\end{equation}
\begin{equation}
\frac{dP_1}{dt} = -\frac{\partial \mathcal{H}_0}{\partial W_1} - \big(\sigma'(\mathbf{z}_1) \odot \bm{\delta}_1\big)\,\mathbf{h}_0^T,
\end{equation}
\end{subequations}
where $\odot$ denotes the componentwise product. A direct differentiation of the loss \eqref{two_layer_loss} confirms that, on the constraint surface,
\begin{equation}
\label{two_layer_gradients}
\frac{\partial L}{\partial W_2} = \bm{\delta}_2\,\mathbf{h}_1^T, \qquad \frac{\partial L}{\partial W_1} = \big(\sigma'(\mathbf{z}_1) \odot \bm{\delta}_1\big)\,\mathbf{h}_0^T,
\end{equation}
which are the familiar backpropagation formulas for a dense network. The Dirac evolution of the momenta therefore contains the exact gradient of the loss, although the Hamiltonian \eqref{two_layer_hamiltonian} depends on the parameters only through $\mathcal{H}_0$. The equations of motion of the parameters themselves are unchanged, $dW_\ell/dt = \partial\mathcal{H}_0/\partial P_\ell$.

\subsection{Regression and classification}
The cost enters equations \eqref{two_layer_errors}--\eqref{two_layer_gradients} only through the output error $\bm{\delta}_2 = \partial c/\partial\mathbf{h}_2$; the matrix $\mathcal{M}^{-1}$, the backward recursion and the brackets \eqref{two_layer_momentum_brackets} are the same for every cost. For least-squares regression with target $\mathbf{y} \in \mathbb{R}^{n_2}$,
\begin{equation}
\label{two_layer_least_squares}
c(\mathbf{h}_2) = \frac{1}{2}\big|\mathbf{h}_2 - \mathbf{y}\big|^2, \qquad \bm{\delta}_2 = \mathbf{h}_2 - \mathbf{y}.
\end{equation}
For classification into $n_2$ classes, the outputs $\mathbf{h}_2$ are interpreted as logits, the predicted class probabilities are $\mathrm{softmax}(\mathbf{h}_2)_k = e^{(\mathbf{h}_2)_k}/\sum_{j} e^{(\mathbf{h}_2)_j}$, and the target $\mathbf{y}$ is a probability vector, $\sum_k y_k = 1$ (a one-hot vector for a labeled example). The cross-entropy cost and its output error are
\begin{equation}
\label{two_layer_cross_entropy}
\begin{gathered}
c(\mathbf{h}_2) = -\sum_{k=1}^{n_2} y_k \log \mathrm{softmax}(\mathbf{h}_2)_k,\\
\bm{\delta}_2 = \mathrm{softmax}(\mathbf{h}_2) - \mathbf{y},
\end{gathered}
\end{equation}
where the normalization $\sum_k y_k = 1$ has been used. In both cases the output error is the difference between the prediction and the target. For two classes, the cross-entropy has a simple physical reading. Writing the label as $y = \pm 1$ and the difference of the two logits as $z = (\mathbf{h}_2)_1 - (\mathbf{h}_2)_2$, the predicted probability of the label is $e^{yz/2}/\big(2\cosh(z/2)\big)$, the Boltzmann distribution, at unit temperature, of an Ising spin $y$ in the magnetic field $z/2$. The cost is then the energy of the observed spin minus the free energy, and the output error with respect to $z$ is $-\frac{1}{2}\big(y - \tanh(z/2)\big)$: the difference between the observed spin and the magnetization predicted by the model, the same moment-matching structure as the learning rule of Boltzmann machines~\cite{ackley1985learning}. Inserting equation \eqref{two_layer_least_squares} or \eqref{two_layer_cross_entropy} into equations \eqref{two_layer_errors} and \eqref{two_layer_gradients} gives the gradients for regression and for classification, respectively.

\subsection{Forward pass}
For the activations, equation \eqref{dirac_activations_network} with $A^{-1}$ from equation \eqref{two_layer_a_matrix} gives
\begin{equation}
\label{two_layer_forward_mode}
\frac{d\mathbf{h}_1}{dt} = \sigma'(\mathbf{z}_1) \odot \big(\dot{W}_1\mathbf{h}_0\big), \qquad
\frac{d\mathbf{h}_2}{dt} = W_2\,\frac{d\mathbf{h}_1}{dt} + \dot{W}_2\,\mathbf{h}_1,
\end{equation}
which is the forward-mode recursion \eqref{forward_mode_recursion}, and coincides with the time derivative of the network \eqref{two_layer_network} along the motion of the parameters.

\subsection{Training}
For heavy-ball momentum, $\mathcal{H}_0$ is the Hamiltonian \eqref{hamiltonian_cov} without $V$, applied to every entry of $W_1$ and $W_2$. Equations \eqref{two_layer_momenta_evolution} and $dW_\ell/dt = \partial\mathcal{H}_0/\partial P_\ell$ then become, in the physical limit, equations \eqref{hamilton_equations_euc} for each weight matrix, which are equivalent to
\begin{equation}
\label{two_layer_ode}
\mu(t)\,\ddot{W}_\ell + \gamma(t)\,\dot{W}_\ell + \frac{\partial L}{\partial W_\ell} = 0, \qquad \ell = 1, 2,
\end{equation}
with the gradients \eqref{two_layer_gradients}. In the gradient-flow limit $\mu \to 0$, the discretization \eqref{discrete_gradient_flow_damping} gives the update
\begin{equation}
\label{two_layer_gd_update}
W_\ell[n+1] = W_\ell[n] - \lambda\,\frac{\partial L}{\partial W_\ell}\Big|_{W[n]}, \qquad \lambda = \frac{h}{\gamma}.
\end{equation}
One step of this update reproduces exactly one iteration of standard training by backpropagation, with each ingredient supplied by the Dirac structure:
\begin{enumerate}
\item the forward pass, $\mathbf{h}_1 = \sigma(W_1\mathbf{h}_0)$ and $\mathbf{h}_2 = W_2\mathbf{h}_1$, is the constraint surface \eqref{two_layer_constraints};
\item the backward pass, equation \eqref{two_layer_errors}, is the action of the block $A^{-T}$ of the inverse constraint matrix;
\item the gradients \eqref{two_layer_gradients} are the brackets \eqref{two_layer_momentum_brackets} of the momenta with the constraints, contracted with the errors;
\item the update \eqref{two_layer_gd_update} is the discretized equation of motion of the optimizer.
\end{enumerate}
For Nesterov's method, the construction is that of Appendix~\ref{appendix_nesterov_lookahead}. Each copy acquires the look-ahead weight matrices $Y_\ell$, with the constraints $Y_\ell - W_\ell - h\bar{V}_\ell \approx 0$, where $\bar{V}_\ell$ is the mean \eqref{lookahead_mean_velocity} of the velocities of $W_\ell$ in the two copies, and the layer constraints \eqref{two_layer_constraints} are imposed with $Y_\ell$ in place of $W_\ell$. Nothing else changes in the network: the matrix \eqref{two_layer_a_matrix} and the errors \eqref{two_layer_errors} are the same functions, evaluated at the look-ahead point, the gradients \eqref{two_layer_gradients} are evaluated at $Y$, and the forward pass \eqref{two_layer_forward_mode} is driven by $\dot{Y}_\ell$ instead of $\dot{W}_\ell$, as in equation \eqref{lookahead_constrained_evolution}. In the physical limit, equation \eqref{two_layer_ode} becomes
\begin{equation}
\label{two_layer_nesterov_ode}
\mu\,\ddot{W}_\ell + \gamma\,\dot{W}_\ell + \frac{\partial L}{\partial W_\ell}\Big|_{W + h\dot{W}} = 0, \qquad \ell = 1, 2,
\end{equation}
which is equation \eqref{nesterov_lookahead_ode} for the network, the continuum counterpart of Nesterov's update \eqref{nesterov},
\begin{equation}
\label{two_layer_nesterov_update}
\begin{gathered}
W_\ell[n+1] = \tilde{W}_\ell[n] - \alpha\,\frac{\partial L}{\partial W_\ell}\Big|_{\tilde{W}[n]},\\
\tilde{W}[n] = W[n] + \beta\big(W[n] - W[n-1]\big).
\end{gathered}
\end{equation}
In both, the forward and the backward pass are evaluated at the extrapolated weights.

\section{Look-ahead constraints of Nesterov's method}
\label{appendix_nesterov_lookahead}

In this appendix we derive the results on Nesterov's method quoted in Sec.~\ref{section_backpropagation}, using the general relations \eqref{total_evolution}--\eqref{constraint_matrix_blocks}.

\emph{Constraints.} Each copy acquires the look-ahead parameters $\mathbf{y}_i$ as coordinates, with conjugate momenta $\bm{\rho}_i$. The Hamiltonian is still \eqref{network_hamiltonian}, and the constraints are
\begin{equation}
\label{lookahead_constraints}
\bm{\psi}_i \equiv \mathbf{y}_i - \mathbf{x}_i - h\,\bar{\mathbf{v}} \approx 0, \qquad \bm{\rho}_i \approx 0,
\end{equation}
together with the layer constraints \eqref{layer_constraints}, in which the parameters $\mathbf{x}_i$ are replaced by $\mathbf{y}_i$. From the Hamiltonian $\ham_0$ of equation \eqref{nesterov_lookahead_hamiltonian}, the mean of the velocities $v^a = \bpartial^a\ham_0 = (\mathbf{v}_1, \mathbf{v}_2)$ is
\begin{equation}
\label{lookahead_mean_velocity}
\bar{\mathbf{v}} = \frac{\mathbf{p}_1 + \mathbf{p}_2}{2\mu} - \frac{\gamma - \dot{\mu}}{4\mu}\,(\mathbf{x}_1 + \mathbf{x}_2).
\end{equation}
We write $\varepsilon_1 = 1$ and $\varepsilon_2 = -1$, so that the bracket of every canonical pair of copy $i$ is $\varepsilon_i$, and denote by $A_i = \partial\bm{\phi}_i/\partial\mathbf{h}_i$ and $B_i = \partial\bm{\phi}_i/\partial\mathbf{y}_i$ the matrices \eqref{a_matrix} and \eqref{phi_h_bracket} evaluated at $\mathbf{y}_i$.

\emph{Constraint algebra.} The nonvanishing brackets between the constraints are
\begin{equation}
\label{lookahead_constraint_brackets}
\begin{aligned}
\Gamma(\bm{\phi}_i, \bm{\pi}_j) &= \varepsilon_i\delta_{ij}A_i, &
\Gamma(\bm{\phi}_i, \bm{\rho}_j) &= \varepsilon_i\delta_{ij}B_i,\\
\Gamma(\bm{\psi}_i, \bm{\rho}_j) &= \varepsilon_i\delta_{ij}\mathbb{I}, &
\Gamma(\bm{\psi}_i, \bm{\psi}_j) &= \frac{h}{2\mu}(\varepsilon_i - \varepsilon_j)\mathbb{I}.
\end{aligned}
\end{equation}
The first is equation \eqref{constraint_brackets_network}. The last follows from $\partial\bm{\psi}_i/\partial\mathbf{x}_k = [-\delta_{ik} + h(\gamma - \dot{\mu})/4\mu]\mathbb{I}$ and $\partial\bm{\psi}_i/\partial\mathbf{p}_k = -(h/2\mu)\mathbb{I}$; it shows that the look-ahead constraints of the two copies do not commute, $\Gamma(\bm{\psi}_1, \bm{\psi}_2) = (h/\mu)\mathbb{I}$, so that the copies can no longer be treated separately. With $Q = (\bm{\psi}_i, \bm{\phi}_i)$ and $P = (\bm{\rho}_i, \bm{\pi}_i)$, the matrix of brackets has the form \eqref{constraint_matrix_blocks}, in which $C$ contains only the block $\Gamma(\bm{\psi}_1, \bm{\psi}_2)$ and $D$ is block diagonal over the copies,
\begin{equation}
\label{lookahead_D_blocks}
D_i = \varepsilon_i\begin{bmatrix} \mathbb{I} & 0\\ B_i & A_i \end{bmatrix}, \quad
D_i^{-1} = \varepsilon_i\begin{bmatrix} \mathbb{I} & 0\\ -A_i^{-1}B_i & A_i^{-1} \end{bmatrix}.
\end{equation}
Since $\det D_i = \pm 1$, the constraints are second class for any architecture, and the block $A$ of the network is unchanged. The inverse matrix, however, acquires the block $D^{-1}CD^{-T}$, which couples the two copies.

\emph{Multipliers.} Let $\mathbf{u}_\psi$, $\mathbf{u}_\phi$, $\mathbf{u}_\rho$ and $\mathbf{u}_\pi$ be the multipliers of the total Hamiltonian. The preservation of $\bm{\rho}_i$ and $\bm{\pi}_i$ in time gives $\mathbf{u}_{\psi,i} + B_i^T\mathbf{u}_{\phi,i} = 0$ and $A_i^T\mathbf{u}_{\phi,i} = -\varepsilon_i\,\partial c/\partial\mathbf{h}_i$, so that
\begin{equation}
\label{lookahead_multipliers}
\mathbf{u}_{\phi,i} = -\varepsilon_i\,\bm{\delta}_i, \qquad \mathbf{u}_{\psi,i} = -\varepsilon_i\,\nabla L(\mathbf{y}_i),
\end{equation}
where $\bm{\delta}_i = A_i^{-T}\partial c/\partial\mathbf{h}_i$ are the errors \eqref{dirac_momenta_network} evaluated at the look-ahead point, and we used $B_i^T\bm{\delta}_i = -\nabla L(\mathbf{y}_i)$. The backward pass is unchanged, and the multiplier of the look-ahead constraint is the gradient of the loss at $\mathbf{y}_i$. The preservation of $\bm{\psi}_i$ and $\bm{\phi}_i$ determines the remaining multipliers. With $\Gamma(\bm{\psi}_i, \mathcal{H}) = -\mathbf{v}_i - h\,\Gamma(\bar{\mathbf{v}}, \mathcal{H}_0)$ and $\sum_j\Gamma(\bm{\psi}_i, \bm{\psi}_j)\,\mathbf{u}_{\psi,j} = (h/\mu)\nabla L(\mathbf{y}_{\bar{\imath}})$, where $\bar{\imath}$ denotes the copy opposite to $i$, we find
\begin{equation}
\label{lookahead_multipliers_momenta}
\begin{aligned}
\varepsilon_i\mathbf{u}_{\rho,i} &= \mathbf{v}_i + h\Big[\Gamma(\bar{\mathbf{v}}, \mathcal{H}_0) + \frac{\partial\bar{\mathbf{v}}}{\partial t}\Big] - \frac{h}{\mu}\nabla L(\mathbf{y}_{\bar{\imath}}),\\
\mathbf{u}_{\pi,i} &= -A_i^{-1}B_i\,\mathbf{u}_{\rho,i}.
\end{aligned}
\end{equation}

\emph{Equations of motion.} Only the constraints $\bm{\psi}_j$ have nonvanishing brackets with the parameters and their momenta, $\Gamma(\mathbf{x}_i, \bm{\psi}_j) = -\varepsilon_i(h/2\mu)\mathbb{I}$ and $\Gamma(\mathbf{p}_i, \bm{\psi}_j) = \varepsilon_i[\delta_{ij} - h(\gamma - \dot{\mu})/4\mu]\mathbb{I}$. Equations \eqref{total_evolution} and \eqref{lookahead_multipliers} then give, with $\mathbf{g}_- \equiv \nabla L(\mathbf{y}_1) - \nabla L(\mathbf{y}_2)$,
\begin{equation}
\label{lookahead_equations_of_motion}
\begin{aligned}
\frac{dx^a}{dt} &= \bpartial^a\mathcal{H}_0 + \frac{h}{2\mu}\,(\mathbf{g}_-, -\mathbf{g}_-),\\
\frac{dp^a}{dt} &= -\partial^a\mathcal{H}_0 - \big(\nabla L(\mathbf{y}_1), \nabla L(\mathbf{y}_2)\big)\\
&\quad + \frac{h(\gamma - \dot{\mu})}{4\mu}\,(\mathbf{g}_-, -\mathbf{g}_-).
\end{aligned}
\end{equation}
These are Hamilton's equations \eqref{hamilton_equations_cov} for the look-ahead Hamiltonian \eqref{nesterov_lookahead_hamiltonian}, in the whole doubled phase space, as one verifies by differentiating equation \eqref{nesterov_lookahead_hamiltonian} with $\partial\bar{\mathbf{v}}/\partial\mathbf{p}_i = \mathbb{I}/2\mu$ and $\partial\bar{\mathbf{v}}/\partial\mathbf{x}_i = -(\gamma - \dot{\mu})\mathbb{I}/4\mu$. Compared with equation \eqref{dirac_momenta_network_cov}, the gradient is evaluated at $\mathbf{y}_i$, and there appear terms proportional to $\mathbf{g}_-$, which vanish in the physical limit, where $\mathbf{y}_1 = \mathbf{y}_2$ and equations \eqref{lookahead_equations_of_motion} reduce to equation \eqref{nesterov_lookahead_ode}. The remaining multipliers give the evolution of the constrained coordinates,
\begin{equation}
\label{lookahead_constrained_evolution}
\begin{aligned}
\frac{d\mathbf{y}_i}{dt} &= \varepsilon_i\mathbf{u}_{\rho,i} = \frac{d\mathbf{x}_i}{dt} + h\,\frac{d\bar{\mathbf{v}}}{dt},\\
\frac{d\mathbf{h}_i}{dt} &= \varepsilon_i\mathbf{u}_{\pi,i} = \frac{\partial\mathbf{h}}{\partial\mathbf{y}}\bigg|_{\mathbf{y}_i}\frac{d\mathbf{y}_i}{dt},
\end{aligned}
\end{equation}
that is, the forward recursion \eqref{forward_mode_recursion} driven by the velocity of the look-ahead point.

\emph{Time dependence.} When $\mu$ or $\gamma$ depend on time, so do the constraints $\bm{\psi}_i$, through $\bar{\mathbf{v}}$: $\partial\bm{\psi}_i/\partial t = -h\,\partial\bar{\mathbf{v}}/\partial t$, and for $\mu = 1$ and $\gamma = 3/t$ we have $\partial\bar{\mathbf{v}}/\partial t = 3(\mathbf{x}_1 + \mathbf{x}_2)/4t^2$. In agreement with the general result of Sec.~\ref{section_constraint_algebra}, this term enters only the multipliers \eqref{lookahead_multipliers_momenta}, and hence only the evolution \eqref{lookahead_constrained_evolution} of $\mathbf{y}_i$ and $\mathbf{h}_i$; the equations of motion \eqref{lookahead_equations_of_motion} are not affected.

\emph{Dirac brackets.} On the constraint surface $\mathbf{y}_i = \mathbf{x}_i + h\bar{\mathbf{v}}$ and $\mathbf{h}_i = \mathbf{h}(\mathbf{y}_i)$, so that the brackets \eqref{network_dirac_jacobian} are replaced by
\begin{equation}
\label{lookahead_dirac_brackets}
\begin{aligned}
\Delta(\mathbf{y}_i, \mathbf{p}_j) &= \Big[\varepsilon_i\delta_{ij} - \varepsilon_j\frac{h(\gamma - \dot{\mu})}{4\mu}\Big]\mathbb{I},\\
\Delta(\mathbf{y}_i, \mathbf{x}_j) &= -\varepsilon_j\frac{h}{2\mu}\,\mathbb{I}, \qquad
\Delta(\mathbf{y}_1, \mathbf{y}_2) = \frac{h}{\mu}\,\mathbb{I},\\
\Delta(\mathbf{h}_i, \,\cdot\,) &= \frac{\partial\mathbf{h}}{\partial\mathbf{y}}\bigg|_{\mathbf{y}_i}\Delta(\mathbf{y}_i, \,\cdot\,).
\end{aligned}
\end{equation}
The activations no longer commute with the parameters and, since $\Delta(\mathbf{y}_1, \mathbf{y}_2)$ does not vanish, those of the two copies no longer commute with each other. For $h \to 0$, equations \eqref{lookahead_dirac_brackets} reduce to equations \eqref{network_dirac_jacobian}.

\bibliography{references}

\end{document}